\documentclass{svproc}
\usepackage{setup}
\usepackage{caption}

\DeclareMathOperator*{\argmax}{arg\,max}

\begin{document}
	\mainmatter
	
	\title{Effects of limited and heterogeneous memory in hidden-action situations}

	\titlerunning{Limited and heterogeneous memory in hidden-action situations}

	\author{Patrick Reinwald \and Stephan Leitner \and Friederike Wall}
	\authorrunning{P. Reinwald et al.}

	\institute{Department of Management Control and Strategic Management,\\ University of Klagenfurt, 9020 Klagenfurt, Austria\\
		\email{\{patrick.reinwald, stephan.leitner, friederike.wall\}@aau.at}
	}

	\maketitle 
\begin{abstract}
	Limited memory of decision-makers is often neglected in economic models, although it is reasonable to assume that it significantly influences the models' outcomes. The hidden-action model introduced by Holmström also includes this assumption. In delegation relationships between a principal and an agent, this model provides the optimal sharing rule for the outcome that optimizes both parties' utilities. This paper introduces an agent-based model of the hidden-action problem that includes limitations in the cognitive capacity of contracting parties. Our analysis mainly focuses on the sensitivity of the principal's and the agent's utilities to the relaxed assumptions. The results indicate that the agent's utility drops with limitations in the principal's cognitive capacity. Also, we find that the agent's cognitive capacity limitations affect neither his nor the principal's utility. Thus, the agent bears all adverse effects resulting from limitations in cognitive capacity.

\end{abstract}
\section{Introduction}
\label{sec:intro}

Over the past decades, the mathematical models and methods used in Economics developed steadily, which finally also led to increasing intelligence of agents and, therefore, to many different implementations of intelligence. Closely related to this large variety of concepts to express intelligence is the concept of (individual) rationality \cite{Thaler.2000,godelier2013rationality}. This concept is concerned with the behavioural aspects of agents. It captures how agents should ideally behave to achieve their objectives, taking into consideration certain (and case-specific) constraints and conditions \cite{godelier2013rationality,simon1972theories}.

An important component of rational decision making is that agents can deal with the complexity of decision problems. This, amongst others, presupposes that agents know how to deal with risk and uncertainty, have access to all relevant information, can identify information as being relevant, and can reason based on the available information. The cognitive capacity required to do so can, at least to some extent, be measured and is usually entitled as general intelligence  or intelligence quotient (IQ) \cite{Thaler.2000,McGrew.2009,Chen.2014,wall2020agent}. 

It is well-known that human cognitive abilities are limited, which is also reflected in certain economic models. Taking limited cognitive capacities into account requires a shift away from the assumptions related to perfectly rational decision makers, which opens the field for various concepts of bounded rationality. However, capturing more natural cognitive abilities in models often requires a change in the modeling (or more generally research) approach, which frequently results in a move from  traditional mathematical methods to simulation-based approaches. This is particularly required since research problems that include decision makers with bounded rationality, due to their innate complexity, usually can not be solved using mathematical methods (i.e., optimization), but often requires simulative approaches \cite{arthur1991designing,Thaler.2000,Axtell.2007,Guerrero.2011,chen2012varieties,Leitner.2015b}. 

We take up on this stream of research and limit the cognitive abilities of the principal and the agent in the hidden-action model introduced by Holmström \cite{Holmstrom.1979} by bounding their cognitive capacity (memory). In the vein of Leitner and Wall \cite{Leitner.2020}, we transfer the hidden-action model into an agent-based model variant using the approach introduced by Guerrero and Axtell \cite{Guerrero.2011}, and Leitner and Behrens \cite{Leitner.2015b}, which enables us to relax selected assumptions to implement the more natural concept of human cognitive ability.

The remainder of this paper is organized as follows: Section \ref{sec:ha-model} elaborates on the hidden-action model originally introduced in Holmström \cite{Holmstrom.1979}. In Section \ref{sec:abm-variant}, we present the agent-based model variant, where we also explain the adjustments to the assumptions necessitated by the changes. Section \ref{sec:results} explain the simulation setup, and introduces and discusses the results.
\section{The hidden-action model}
\label{sec:ha-model}
The hidden-action model, which was amongst others introduced by Holmström \cite{Holmstrom.1979}, is a single-period model that describes a situation in which a principal assigns a task to an agent.  For this:
\begin{itemize}
	\item The principal designs a contract upon the task to carry out, also including an incentive scheme, i.e., a sharing rule that fixes how the task's outcome will be shared between the principal and the agent.
	\item The principal offers the contract to the agent. 
	\item The agent decides whether to accept the contract or not. If the agent agrees on the conditions stated in the contract, he decides how much effort (also referred to as action) he wants to make in fulfilling the specified task.
	\item The agent's chosen effort level generates, together with an exogenous factor (also referred to as environmental factor), the outcome of the task.
	\item  Carrying out the task leads to disutility for the agent, depending on his effort level.
	\item The principal is not able to observe the action of the agent, so it is hidden. This results in the situation that only the outcome serves as basis for the incentive scheme.
	\item The model assumes that the principal is risk-neutral and the agent is risk-averse. Both are individual utility maximizers.
\end{itemize}
The utility function $U_P(x,s)=x-s(x)$ characterizes the principal, whereby $x$ denotes the generated outcome, and $s=s(x)$ is the function of the sharing rule. The outcome is calculated  as $x=f(a,\theta)$, which is a function of the agent's chosen effort level $a$ and the exogenous factor $\theta$. The agent is characterized by the utility function $U_A(s,a)=V(s(x))-G(a)$, where $V(s)$ represents the utility generated from compensation and $G(a)$ is the function for the disutility generated from exerting effort. The optimization problem to generate the optimal solution is formulated as follows:
\begin{eqnarray}
\underset{s(x),a}\max E(U_P(x-s(x)))\\
\label{eq:parti}
s.t~ E\{U_A{(s(x),a)\} \geq \bar{U}}\\
\label{eq:incenti}
a \in \underset{a' \in A}{\argmax} \: E\{U_A(s(x),a')\} ~, 
\end{eqnarray}
where Eqs. (\ref{eq:parti}) and (\ref{eq:incenti}) are constraints that have to be considered by the principal. The notation "arg max" represents the set of all arguments that maximizes the objective function that follows. Equation \ref{eq:parti} is referred to as the participation constraint, which ensures that the agent accepts the offered contract by ensuring him the minimum utility $\bar{U}$. This minimum utility is also referred to as reservation utility and represents the agent's best outside option. Equation \ref{eq:incenti} is known as the incentive compatibility constraint and aligns the agent's objective (maximize his utility) with the principal's objective. This constraint affects the agent's choice of effort level $a$ \cite{Eisenhardt.1989,BernardCaillaud.2000,Lambert.2001,Reinwald.2020,Leitner.2020}. In Table \ref{tab:1}, the notation used in this section is summarized.
\begin{table}[h]
	\centering
	\caption{Notation for the hidden-action model}
	\label{tab:1}       % Give a unique label
	\begin{tabular}{ll}
		\hline\noalign{\smallskip}
		Description& Parameter \\
		\noalign{\smallskip}\hline\noalign{\smallskip}
		Principal's utility function & $ U_P(x-s(x)) $ \\
		Agent's utility function & $ U_A(s(x),a) $\\
		Agent's utility from compensation & $ V(s(x)) $\\
		Agent's disutility from effort &$  G(a) $\\
		Agent's reservation utility & $ \bar{U} $\\
		Agent's share of outcome & $ s(x)=x*p $\\
		outcome & $ x=x(a,\theta) $\\
		Premium parameter & $ p $\\
		Effort level & $ a $\\
		Set of all feasible actions & $ A $\\
		Random state of nature & $ \theta $\\
		\noalign{\smallskip}\hline
	\end{tabular}
\end{table}
\section{Agent-based model variant}
\label{sec:abm-variant}
\begin{figure}
	\label{flow}
	\centering
	\includegraphics[scale=0.60]{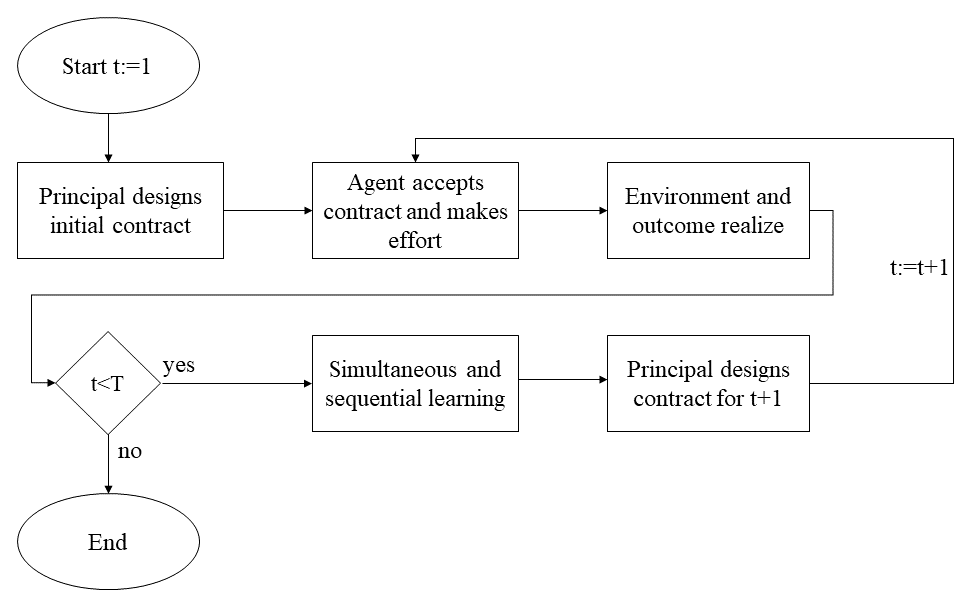}
	\caption{Flow diagram }
	\label{fig:flow}
\end{figure}

The hidden-action model has a set of (rather restrictive) assumptions incorporated in order to make it mathematically tractable \cite{Axtell.2007}. Overviews of the most important assumptions are provided by Axtell \cite{Axtell.2007} and Müller \cite{Mueller.1995}. For this paper, we relax the assumptions of information symmetry regarding the environmental factor as follows:
\begin{itemize}
	\item The principal and the agent no longer can access all information about the exogenous factor at the beginning of the sequence of events. They only know that it follows a Normal Distribution.
	\item The principal and the agent are endowed with learning capabilities so that they are both able to individually learn about the exogenous factor over time. We refer to this learning model as simultaneous and sequential learning.
	\item The principal and the agent have a defined memory to individually store their gathered information about the exogenous factor.
\end{itemize}
As a consequence of these adaptions, the principal and the agent are no longer able to find the optimal solution immediately, which in exchange enables the introduction of a learning mechanism. Furthermore, the adaptation of the assumptions and the resulting change in the research methodology leads to the switch from a single-period model to a multi-period model. The timesteps of the multi-period model are indicated by $t=1,...,T$. An overview of the sequence of events is provided in Figure \ref{fig:flow}, and the notation of the model is summarized in Table \ref{tab:abm}.

The risk-neutral principal is characterized by the utility function
\begin{equation}
\label{utilp_abm}
U_P(x_t,s(x_t))=x_t-s(x_t) ,
\end{equation}
where $xt=a_t+\theta_t$ represents the outcome and $s(x_t)=x_t*p_t$ denotes the agent's compensation in timestep $t$, $a_t$ is the agent's action, $\theta_t$ stands for normally distributed exogenous factor ($\theta_t \raisebox{-0.9ex}{\~{}} N(\mu,\sigma)$), and $p_t \in [0,1]$ stands for the premium parameter of the sharing-rule which is used to compute the agent's compensation in $t$. The risk-averse agent is characterized by the utility function 
\begin{equation}
\label{utila_abm}
U_A(s(x_t),a_t)= \overbrace{\dfrac{1-e^{-\eta*s(x_t)}}{\eta}}^{V(s(x_t)))}-\overbrace{\dfrac{a_t^2}{2}}^{G(a_t)},
\end{equation}
where $\eta$ represents the agent's Arrow-Pratt measure of risk-aversion \cite{arrow1973role}. The relaxation of the assumption mentioned above requires to set boundaries for the space of effort-levels $A$ for every $t$, so that the existence of solutions is assured. This is necessary since the set of all feasible solutions, $A_t$ ($A_t \subset A$), strongly depends on the expectation of the environment and, therefore, can change for every timestep $t$. According to Holmström \cite{Holmstrom.1979}, we identify the lower boundary by the participation constraint and the upper boundary by the incentive compatibility constraint. It is important to notice that both boundaries include an expectation about the exogenous factor $\theta$ (see Eqs. \ref{theta_abmP} and \ref{theta_abmA}) and, therefore, the value of the boundaries might change in every $t$. 

In every timestep $t= 2,...,T$, the principal randomly discovers 2 alternative effort-levels in the search space $A_t$, which together with the action incited in the previous period, $\tilde{a}_t$, serve as candidates for the action in the next period, $\tilde{a}_{t+1}$, and evaluates them with respect to increases in expected utility (based on the utility function in Eq. \ref{utilp_abm}). The discovered effort-levels are modeled to be uniformly distributed in the search space. Please notice that the evaluation of the effort-levels includes the principal's expectation about the exogenous factor in $t$. The principal, as well as the agent, are able to remember previous observations or estimations (cf. Eqs. \ref{theta_abmP} and \ref{theta_abmA}). The retrievable values (defined by the memory $m_P$ and $m_A$ for the principal and the agent, respectively) are subsequently used to compute the expectation of the exogenous factor. For the principal this expectations is computed as follows:
\begin{equation}
\hat{\theta}_{Pt}=
\begin{cases}
\frac{1}{t-1}~ \sum\limits_{n=1}^{n=t-1}{\tilde{\theta}}_{n} &\text{if~} m_P=\infty~,\\
\frac{1}{m_P} \sum\limits_{\substack{\forall t \leq m_P: n=1 \\ \forall t > m_P: n = t-m_P }}^{n=t-1}{\tilde{\theta}}_{n} &\text{if~} m_P<\infty~,\\
\end{cases} 
\label{theta_abmP}
\end{equation} 
The expected outcome (using the value-maximizing effort-level from the principal's point of view, $\tilde{a}_t$) in $t$, thus, be formalized by $\tilde{x}_{Pt}=\tilde{a}_t+\hat{\theta}_{Pt}$. Next,the principal computes the corresponding premium parameter according to 
\begin{equation}
p_t = \max_{p=[0,1]} U_P(\tilde{x}_{Pt},s(\tilde{x}_{Pt}))~.
\end{equation}
The principal offers the contract (which, amongst others, includes $p_t$) to the agent who decides  whether to accept the contract or not. In case the agent accepts the contract, he selects an effort level $a_t = \max_{a \in A_t} U_A(s(x_t),a)$, where $A_t$ represents the space of all feasible effort-levels. As the agent is able to observe the realization of exogenous factors which represents environmental uncertainty, the expectation of the exogenous factor in $t$ from the agent's point of view is computed as follows:
\begin{equation}
\hat{\theta}_{At}=
\begin{cases}
\frac{1}{t-1}~ \sum\limits_{n=1}^{n=t-1}{\theta}_{n} &\text{if~} m_A=\infty~,\\
\frac{1}{m_A} \sum\limits_{\substack{\forall t \leq m_A: n=1 \\ \forall t > m_A: n = t-m_A }}^{n=t-1}{\theta}_{n} &\text{if~} m_A<\infty~,\\
\end{cases} 
\label{theta_abmA}
\end{equation} 
Next, the exogenous factor, $\theta_t$, and the outcome $x_t$, realize. The agent can observe $x_t$ and $\theta_t$ and memorizes $\theta_t$. The principal can only observe $x_t$, estimates the exogenous factor according to
\begin{equation}
\label{calcThetaPrincipal}
\tilde{\theta}_t=x_t- \tilde{a}_t ~,
\end{equation}
and memorizes $\tilde{\theta}_t$.\footnote{Please notice that, as long as only one piece of information is unavailable, $\tilde{a}_t$ and $a_t$ perfectly coincide, and, thus, the principal can estimate the realization of the exogenous factor without error.}  Finally, the utilities for the principal and the agent realize, and $\tilde{a}_t$ is carried over to period $t+1$ as status-quo effort level. This sequence is repeated $T$ times.
\begin{table}[H]
	\centering
	\caption{Notation for the agent-based model variant}
	\label{tab:abm}       % Give a unique label
	\begin{tabular}{lll}
		\hline\noalign{\smallskip}
		Description& Parameter \\
		\noalign{\smallskip}\hline\noalign{\smallskip}
		Timesteps & $ t $ \\
		Principal's utility & $ U_P $ \\
		Agent's utility & $ U_A $ \\
		Agent's Arrow-Pratt measure of risk-aversion & $ \eta $ \\
		Agent's share of outcome in $t$ & $ s(x_t)=x_t*p_t $ \\
		Outcome in $t$ & $x_t=a_t+\theta_t$ \\
		Principal's expected outcome & $ \tilde{x}_{Pt} $\\
		Premium parameter in $t$ & $p_t$\\
		Agent's chosen effort level in $t$ & $ a_t $\\
		Induced effort level by the principal in $t$ & $ \tilde{a}_t $\\
		Set of all actions & $ A $ \\
		Set of all feasible actions in $t$ & $ A_t $ \\
		Exogenous (environment) variable in $t$ & $\theta_t$\\
		Principal's estimation of the realized exogenous factor in $t$ & $\tilde{\theta_t}$ \\
		Principal's memory in periods & $m_P$\\
		Agent's memory in periods & $m_A$\\
		Averaged expected exogenous factor of the principal & $ \hat{\theta}_{Pt} $\\
		Averaged expected exogenous factor of the agent & $ \hat{\theta}_{At} $\\
		
		\noalign{\smallskip}\hline
	\end{tabular}
\end{table}
\section{Results}
\label{sec:results}
\subsection{Parameterization}
\label{sec:para}
For this paper, the analysis puts special emphasis on the principal's and the agent's memory (cognitive capacity), and the distribution of the environment on the utility of both, the principal and the agent. All other parameters are kept constant during the simulation runs. The parameterization is summarized in Tab. \ref{tab:key_para}.

The parameterization allows distinguishing two levels of cognitive capacity by the variable $m_P$ for the principal, and $m_A$ for the agent. They have either limited cognitive capacity with a memory of $3$ periods or unlimited cognitive capacity with a memory of $T-1$ periods, whereby all four combinations of cognitive capacity are simulated.  The environment in which an organization operates can be characterized by $\sigma$, whereby an organization can either act in relatively stable ($\sigma=0.05x^*)$, moderately stable ($\sigma=0.25x^*)$, or unstable ($\sigma=0.45x^*$) environment. The characterization of both parties (the principal and the agent) and organizations leads to a total number of 12 scenarios. 
\begin{table}[h]
	\centering
	\caption{Key parameters}
	\label{tab:key_para}       % Give a unique label
	\begin{tabular}{lll}
		\hline\noalign{\smallskip}
		Parameter & Notation & Values \\
		\noalign{\smallskip}\hline\noalign{\smallskip}
		Principal's memory & $ m_P $ & $\{3,\infty\}$\\
		Agent's memory & $ m_A $  & $\{3,\infty\}$\\
		Exogenous factor: standard deviation & $ \sigma $ & $\{0.05x^*, 0.45x^*\}$\\
		Exogenous factor: mean & $ \mu $ & $ 0$\\
		Agent's Arrow-Pratt measure & $ \eta $ & $0.5$\\
		\noalign{\smallskip}\hline
	\end{tabular}
\end{table}
\subsection{Simulations and reported performance measure}
In total, there are 12 investigated scenarios (as described in Sec. \ref{sec:para}). For each scenario we simulated $R=700$ paths\footnote{This number of paths appears to be appropriate as a relatively stable coefficient of variation can be observed (cf. \cite{lorscheid2012opening})} and in each path we simulated $T=20$ timesteps. In each timestep the principal is able to adapt the parameterization of the incentive scheme, and the agent is able to change the effort he makes. 
\subsection{Results and Discussion}
\begin{figure}[H]
	\centering
	\includegraphics[scale=0.23]{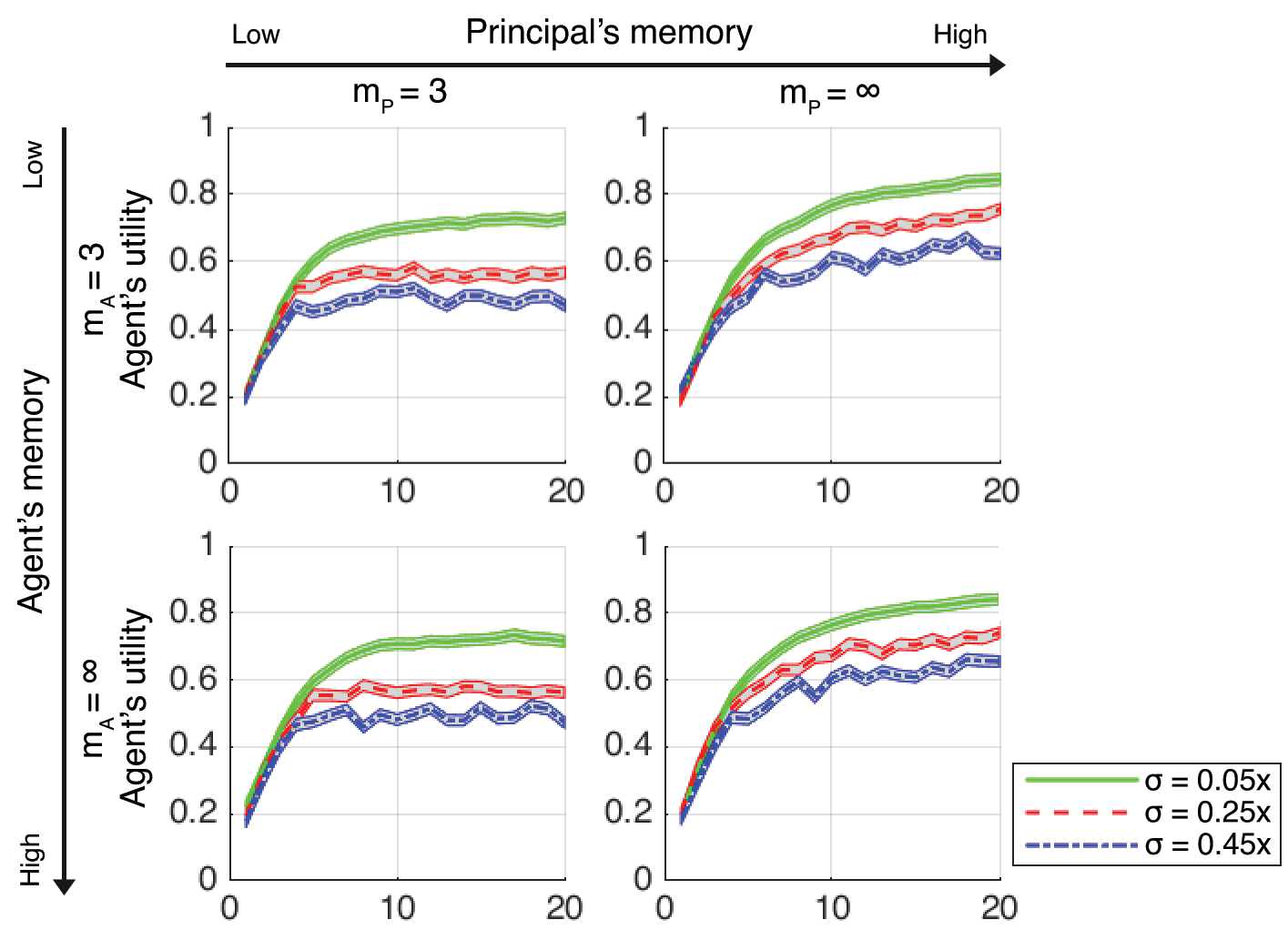}
	\caption{Agent's utility for all scenarios. Shaded areas represent confidence intervals at the 99\%-level}
	\label{fig:UA}
\end{figure}

\begin{figure}[H]
	\centering
	\includegraphics[scale=0.23]{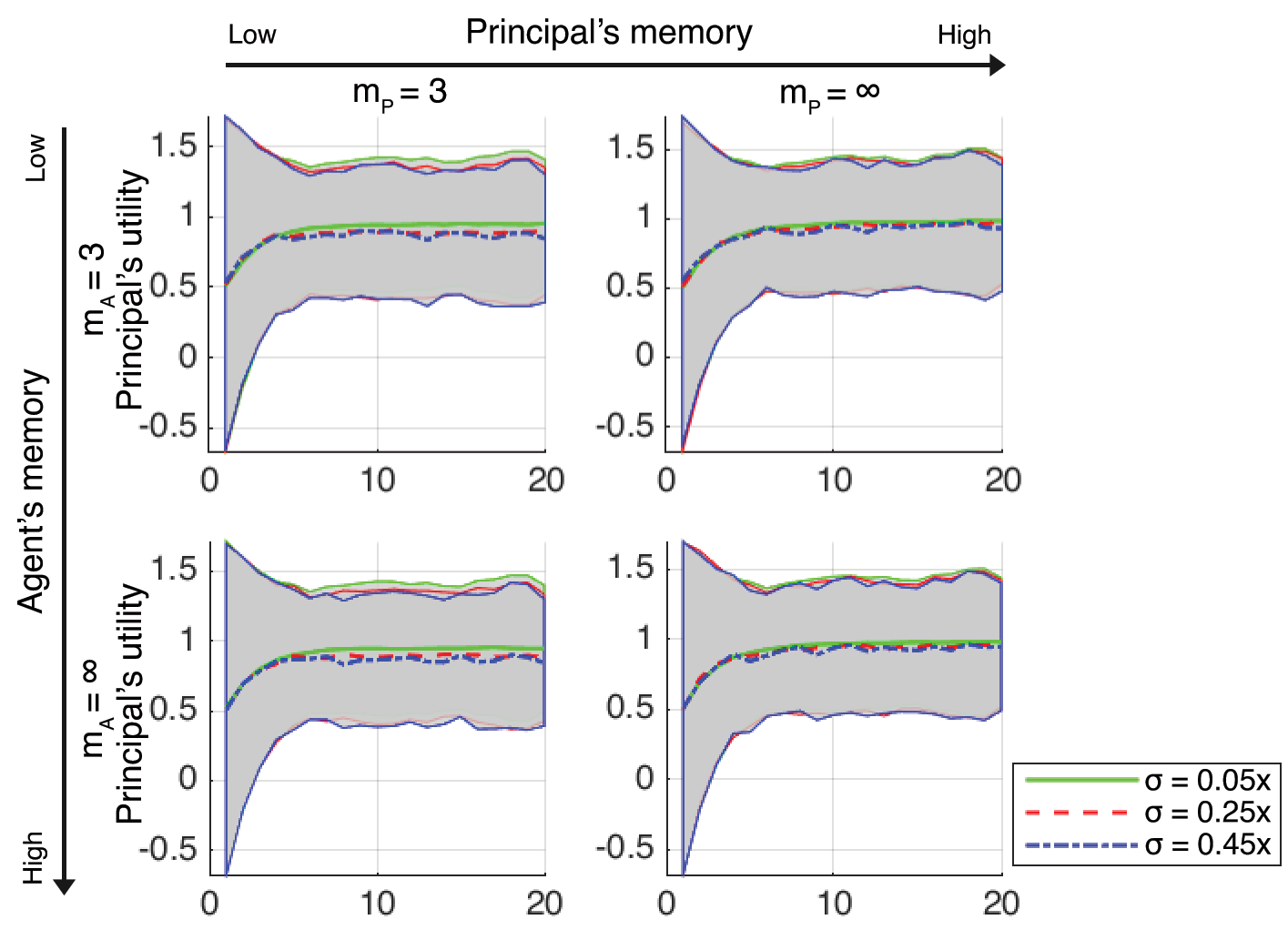}
	\caption{Principal's utility for all scenarios. Shaded areas represent confidence intervals at the 99\%-level}
	\label{fig:UP}
\end{figure}

The results for the agent's utility are shown in Fig. \ref{fig:UA}, while Fig. \ref{fig:UP} displays the results for the principal's utility. Each figure consists of four subplots, representing different combinations of memory (cognitive capacity). In every subplot three lines are plotted, each representing a different environmental uncertainty (represented through different standard deviations $\sigma= \{0.05x^*, 0.25x^*, 0.45x^*\}$).

For the agent's utility we can see in all scenarios, that a higher extent in environmental uncertainty significantly decreases his utility. This may be a result of the overall lower outcome of the task, which was already shown in Reinwald et al. \cite{Reinwald.2020}, and is in line with the existing literature on environmental uncertainty \cite{milliken1987three,chen2005impacts} and environmental dynamism \cite{aldrich2008organizations}. These papers suggests that more uncertain and dynamic environments increase the difficulty of organizational decision-making and, consequently, is a significant determinant of organizational performance.

By comparing the different subplots in Fig. \ref{fig:UA}, we can see that the cognitive capacity of the principal (memory $m_P$) has a significant influence on the agent's utility in a way that an increase in the principal's memory also leads to an increase in utility for the agent for all levels of environmental uncertainty. This leads to the suggestion that an increase in cognitive capacity of the principal reduces the uncertainty of the environment for her, and, therefore, results in a better contract, which increases the agent's utility. Regarding the agent's cognitive capacity (memory $m_A$), no effects can be observed.

For the principal's utility, we can not see any significant differences, neither by comparing subplots (different cognitive capacities) nor by comparing the plotted lines within a subplot (different environmental situations). This leads to the conclusion that neither the environmental uncertainty nor changes in the cognitive capacity influence the principal's utility. When compared to the results presented in Reinwald et al. \cite{Reinwald.2020} regarding the task's performance, where we can see that environmental uncertainty has a negative effect on performance and the principal's cognitive capacity has a positive effect, two conclusions arise. The first is, that the risk of environmental uncertainty is solely carried by the agent, and the second is, that the principal has no intention to improve her cognitive capacity.

%\vspace{-1.6cm}

\section{Conclusion}
\label{sec:conclusion}
We transferred the hidden-action model of Holmström into an agent-based variant, and we relaxed the assumption of information symmetry about the environmental factor. We find that the agent's utility is highly affected by both the level of environmental uncertainty and the cognitive capacity of the principal. In contrast, it was observed that the cognitive capacity, in terms of memory, has no effect, neither on the utility of the agent nor on the utility of the principal. Further, the utility of the principal was not affected by any change of the scenarios. 

Our research has certain limitations which call for further research efforts. First, some assumptions incorporated in the hidden-action model are carried over. These assumptions cover, among others, the principal's knowledge of the agent's characteristics, the individual utility-maximizing behaviour, or the ability to perform all actions without errors or biases. Second, there is only one kind of contract (incentive scheme) available. Third, there are no learning-curve effects, although the agent repeatedly carries out the same task. Further research might want to deeper investigate the effects shown in this paper and additionally relax further assumptions incorporated in the hidden-action model. . 

\section*{Acknowledgments}
This work was supported by funds of the Oesterreichische Nationalbank (Austrian Central Bank, Anniversary Fund, project number: 17930).

\bibliographystyle{splncs03}
\bibliography{bibio}
\end{document}